\documentclass[12pt,a4paper]{article}
\usepackage{a4wide}
\usepackage{makeidx}
\usepackage{latexsym}
\usepackage{epsf}
\usepackage{amssymb}
\makeindex

\begin{document}
\def\be{\begin{equation}}
\def\ee{\end{equation}}
\def\bea{\begin{eqnarray}}
\def\eea{\end{eqnarray}}

\def\pd{\partial}
\def\a{\alpha}
\def\b{\beta}
\def\g{\gamma}
\def\d{\delta}
\def\m{\mu}
\def\n{\nu}
\def\t{\tau} 
\def\l{\lambda}
\def\s{\sigma}
\def\e{\epsilon}
\def\scri{\mathcal{J}}
\def\cM{\mathcal{M}}
\def\tcM{\tilde{\mathcal{M}}}
\def\RR{\mathbb{R}}
\def\CC{\mathbb{C}}

\hyphenation{re-pa-ra-me-tri-za-tion}
\hyphenation{trans-for-ma-tions}


\begin{flushright}
IFT-UAM/CSIC-00-33\\
hep-th/0104211\\
\end{flushright}

\vspace{1cm}

\begin{center}

{\bf\Large A Comment on the K-theory Meaning of the Topology of Gauge Fixing}

\vspace{.5cm}

{\bf C\'esar G\'omez}
\footnote{E-mail: {\tt cesar.gomez@uam.es  }}\\
\vspace{.3cm}

\vskip 0.4cm

\
Instituto de F\'{\i}sica Te\'orica, C-XVI,
  Universidad Aut\'onoma de Madrid \\
 { E-28049-Madrid, Spain}\footnote{Unidad de Investigaci\'on Asociada
  al Centro de F\'{\i}sica Miguel Catal\'an (C.S.I.C.)}

\vskip 1cm


{\bf Abstract}

\end{center}
Based on the recently discovered K-theory description of D-branes in
string theory a K-theory interpretation of the topology of gauge conditions
for four dimensional gauge theories, in
particular 't Hooft abelian projection, is presented. The interpretation
of the dyon effect in terms of $K^{-1}$ is also discussed. In the context of
type IIA strings, D-p branes are also interpreted as gauge fixing singularities.


\begin{quote}

\end{quote}


\newpage

\setcounter{page}{1}
\setcounter{footnote}{1}

\section{Topology of Gauge Fixing}
In reference \cite{thooft} a ghost free unitary gauge for non abelian gauge theories
was presented. The idea consist in using some
extra scalar field $X$ transforming in the adjoint representation and to fix the
non abelian part of the gauge
by impossing $X$ to be diagonal
\be\label{1}
X=
\left(
\begin{array}{ccc}
\lambda_{1}&~&~\\
~&\ddots&~\\
~&~&\lambda_{N}\\
\end{array}
\right)
\ee
The subgroup of gauge transformations under which (\ref{1}) is invariant 
is just the maximal
abelian subgroup, $U(1)^{N}$ in the case of $U(N)$ gauge group. 
The spectrum of physical degrees of freedom in this gauge
are $N$ massless photons, $\frac{1}{2}N(N-1)$ massive charged vector fields - 
corresponding to the
off diagonal part of the $A_{\mu}$ gauge matrix - 
and the $N$ scalar fields ${\lambda}_{i}(x)$
defined by (\ref{1}). 
In addition to these degrees of freedom we have topological pointlike
defects -to be identified with magnetic monopoles- corresponding to the singularities
of the gauge fixing (\ref{1}) when two eigenvalues coincide.

The topological description of these gauge fixing singularities is as follows. The
condition for two eigenvalues to coincide fixes a codimension three singularity in
space-time i.e a pointlike particle if we work in four dimensions. In the vicinity
of the singularity the 
matrix $X$, prior to diagonalization, can be written like \cite{thooft}

\be\label{2}
X=
\left(
\begin{array}{c|c|c}
D_{1}&0&0\\\hline
0&
\begin{tabular}{cc}
$\lambda + \epsilon_{3}$&$\epsilon_{1}-i \epsilon_{2}$\\
$\epsilon_{1}+i \epsilon_{2}$&$\lambda-\epsilon_{3}$\\
\end{tabular}
&0\\\hline
0&0&D_{2}\\
\end{array}
\right)
\ee
where $D_{1}$ and $D_{2}$ are diagonal matrices.
If we concentrate our attention on the $2 \times 2$ small matrix in (\ref{2}) we can write

\be
X=\lambda I + \sum_{i=1}^{3}\sigma_{i}x_{i}
\ee
for $\sigma_{i}$ the Pauli matrices. More simply $X=\sum_{i=1}^{3}\sigma_{i}x_{i}$ for 
$\epsilon_{i}(x)=x_{i}$ in
the vicinity of the singular point defined by $\epsilon(x=0)=0$. Now we can easily
associate the field $X=\sum_{i=1}^{3} \sigma_{i}x_{i}$ on $R^{3}$ 
with a magnetic monopole. 
Let us
consider the sphere $S^{2}$ embeded in $R^{3}$. For $x$ lying on $S^{2}$ we have

\be
X^{2}(x) = I
\ee
Therefore we can define two non trivial line bundles $L_{\pm}$ on $S^{2}$
with the fiber defined by the eigenspaces
in $C^{2}$

\be\label{eigen}
X v = {\pm}v
\ee

The first Chern class of these line bundles is the magnetic charge of the monopole.
The physical meaning of these ghost free unitary gauges is based on the following 
general argument. Let us consider for simplicity an abelian Higgs model. In perturbation
theory the Higgs phase is usually defined by the condition

\be\label{higgs}
<\phi> = F 
\ee
for some non vanishing vacuum expectation value $F$. The problem with (\ref{higgs}) as
a non perturbative way to define the Higgs phase, is of course that $\phi$ 
is not a gauge invariant
quantity. We can always choose a gauge of type $Re(\phi) > 0$, $Im(\phi) = 0$ in
such a way that $<\phi>$ is non vanishing regardless what condensation takes place. This
ghost free gauge condition has however a problem, namely the existence of singular
points where $Re(\phi) = 0$. In fact around this point the gauge fixing $Im(\phi) = 0$
determines a singular gauge transformation with non vanishing vorticity. This is the
topological ancestor of the well known Nielsen-Olesen vortices in abelian superconductivity.
The important point here is that the existence of this topological defect is independent
on the concrete Higgs condensation taking place in the system. It only depends on
the topology of the gauge fixing. The difference between the Higgs and the Coulomb
phase is that in the Higgs phase these vortices would be
also the macroscopic degrees of freedom, while in the Coulomb phase they will condense 
into the vacuum. The same type of 
argument for the magnetic monopoles described above will
differenciate the Higgs phase from the confinement 
phase in pure non abelian gauge theories.
The key aspect of 't Hooft definition of microscopic variables 
is that the topological charges
associated with the topology of the gauge fixing are stable with respect to the
concrete condensation taking place. So for instance in the confinement phase these
magnetically charged objects exist like in the Higgs phase, the difference is 
that in the confinement phase they 
fill the vacuum creating an electric superconductor. 
In this note we will offer a K-theory interpretation of this gauge fixing topology. In
order to gain some intuition let us before describe 
very briefly the K-theory interpretation
of D-branes in type II string theory from the point of view of gauge fixing topology.

\section{D-branes and K-theory}
The connection of D-brane charges and RR fields with K-theory has been worked out 
in a series of recent works 
\cite{mm} \cite{witten1} \cite{horova} \cite{mw} \cite{witten2}. 
Let us start considering the description
of D-$p$ branes ( $p$ even) in type IIA string theory \cite{horova}. We 
will start with a set of
$N$ unstable filling 9-branes. The low energy spectrum 
on the 9-brane world volume is (up
to fermionic fields) a $U(N)$ gauge field $A$ 
and an open tachyon $T$ transforming in the 
adjoint representation. Based on the approach
presented in the previous section it
is very natural to use the tachyon field $T$ 
to define a ghost free unitary gauge. If we are working in critical
ten dimensional space-time, gauge fixing singularities 
where two eigenvalues of $T$ coincide, are codimension
three, i.e the natural candidates for D-6 branes. Notice that we are not saying
anything on the particular form of the tachyon potential. 
We are just considering, as we did in
the previous section, a gauge where $T$ is diagonal and we look into the singularities
when two eigenvalues coincide. 

The $T$ field in the vicinity of the singularity is just
given by the matrix (\ref{2}), which in particular means that on the $R^{3}$ transversal
directions we have

\be\label{T}
T(x) = \sum_{i=1}^{3}x_{i}\sigma_{i}
\ee
This is in fact the definition of the tachyon field  \cite{horova}, \cite{witten1} around
the solitonic D-6 brane. In order to define the charge of the D-6 brane we simply repeat
what we did in previous section in order to define a
non trivial line bundle on $S^{2}$, but this time replacing the field  
$X$ by the tachyon field $T$.
In this sense we notice that the D-6 brane can be simply interpreted, independently of
the particular form of the tachyon potential, as a topological defect of the gauge fixing
in ten dimensional gauge theory, in perfect analogy with the topological defects in
the four dimensional abelian projection. Singularities of higher codimension $2k+1$ will
correspond to $2k$ equal eigenvalues. The only difference is that in this case we
should use -in order to define $T$ in the vicinity of the singularity- the Gamma matrices
$\Gamma_{i}, i= 1, ...,2k+1$. The corresponding non trivial vector bundles on $S^{2k}$ are
defined in a similar way with the fiber given 
by the eigenspaces of $T$ ( see (\ref{eigen})). The dimension 
of these bundles is $2^{k-1}$.

It was argued in \cite{witten1} and developped in \cite{horova} that type IIA D-brane
charges should be interpreted in terms of the higher K-group $K^{-1}(X)$. The definition 
of $K^{-1}(X)$ is 

\be\label{k}
K^{-1}(X) = \tilde{K} (SX)
\ee
where $\tilde{K}(SX)$ is the reduced K group of $SX$ 
and where $SX$ is the reduced suspension of $X$
\footnote{The reduced $\tilde{K}$ group of $X$ is defined as follows. 
We first notice that the
K group of a single point is just equal to $Z$. In fact the vector bundles on a point are
simply classified by their dimension which is a natural number, the corresponding
K group is therefore simply the set of integer numbers $Z$. 
If we now choose a point $p$
in $X$ we can map $K(X)$ into $K(p)$, the kernel 
of this map is the reduced $\tilde{K}$ group
of $X$. The reduced suspension $SX$ of a compact space $X$ is defined as $S^{1}\wedge X$,
where $X \wedge Y = X \times Y / X \times y_{0} \bigcup Y \times x_{0}$
for $x_{0}$ and $y_{0}$ base points of $X$ and $Y$ respectively.
The quotient space $X/Y$ for $X$ a compact space is defined
by identification of $Y$ to a single point. For instance $SS^{n}=S^{n+1}$.
See \cite{atiya}, \cite{karoubi} for mathematical details.}. The relation
of type IIA string theory and M-theory together with (\ref{k}) strongly suggest a
M-theory interpretation of $K^{-1}(X)$. In type IIB the
charges of stable D-$p$ branes with even codimension $2k$ are related to the group
$\tilde{K}(S^{2k})$. Elements in $\tilde{K}(S^{2k})$ 
are one to one related with the homotopy
group $\pi_{2k-1}(U(2^{k-1}))$. The M-theory version of type IIA naturally leads
to describe
stable D-$p$ branes of codimension
$2k+1$ in terms of the higher group $K^{-1}(S^{2k+1})$ or equivalently
$\tilde{K}(S^{2k+2})$ instead of $\tilde{K}(S^{2k+1})$, which,
by Bott periodicity theorem, is zero.
This means that the relevant homotopy group should be $\pi_{2k+1}(U(2^{k}))$
which is one to one related with $\tilde{K}(S^{2k+2})$.

The description in terms of homotopy groups is directly connected with
Sen's physical picture \cite{sen} 
of D-branes as solitons resulting from tachyon condensation. In
the case of type IIB $U(2^{k-1})$ is assumed to be isomorphic with the vacuum manifold
for a non vanishing expectation value of the tachyon. In type IIA case the equality
$\pi_{2k+1}(U(2^{k}))= \pi_{2k} (\frac{U(2^{k})}{U(2^{k-1}) \times U(2^{k-1})})$ again
relates $\tilde{K}(S^{2k+2})$ with the assumed vacuum 
manifold $\frac{U(2^{k})}{U(2^{k-1}) \times U(2^{k-1})}$
for the corresponding tachyon.

Coming back to $K^{-1}(S^{2k+1})$ we can easily relate this group 
to $K^{-1}(B^{2k+1},S^{2k})$. This follows from the representation of
$S^{2k+1}$ as the quotient space $B^{2k+1}/S^{2k}$ \footnote{ Here we have used 
the relation $K(B^{n}, S^{n-1}) = K(B^{n}/S^{n-1})$.}. In summary, we
observe how the type IIA D-brane charges
with codimension $2k+1$, are in $K^{-1}(B^{2k+1},S^{2k})$ \cite{horova}.

It is important to stress that from the point of view of gauge fixing the existence
of D-$p$ branes, with $p$ even,  
in type IIA does not depend on the particular tachyon condensation
taking place. They exist as part of the physical 
degrees of freedom of the $U(N)$ gauge theory
associated to filling 9 branes.

\section{K-theory and Gauge Fixing}
Next we would like to describe the topology of gauge fixing using K-theory. As
observed in \cite{horova} there is a different characterization of $K^{-1}$
that does not use any extra dimension \cite{karoubi}. The first ingredient is that
the $K^{-1}$ group of the semigroup of trivial vector bundles on a compact space $X$ is
isomorphic to the $K^{-1}$ group of the semigroup of locally trivial vector bundles
on $X$. This allows us to reduce the discussion to trivial vector bundles $I^{n}$ where
$n$ denotes the dimension. Elements in $K^{-1}$ are now related to pairs $(I^{n},\alpha)$
where $\alpha$ is an automorphism of $I^{n}$. More precisely we define elementary
pairs $(I^{n}, \alpha)$ if the corresponding automorphism
$\alpha$ is homotopic to the identity within the automorphisms
of $I^{n}$. We say that two pairs $(I^{n},\alpha)$ and $(I^{m},\beta)$ are equivalent
if there exist two elementary pairs $(I^{p},\gamma_{1})$ and $(I^{q},\gamma_{2})$
such that  $(I^{n} \bigoplus I^{p},\alpha \bigoplus \gamma_{1})$ 
is isomorphic to $(I^{m} \bigoplus I^{q},\beta \bigoplus \gamma_{2})$
where $\bigoplus$ represents the direct sum. In plain terms the elements in $K^{-1}$ are
going to be one to one related with the homotopy class of the automorphism $\alpha$
\footnote{ More precisely the isomorphism between $ \lbrack X, U(\infty) \rbrack $ and
$K^{-1}(X)$ maps the homotopy class in $ \lbrack X, U(\infty) \rbrack $ to the automorphism
$\alpha$ defining the corresponding element in $K^{-1}(X)$ . See theorem 3.17 in \cite{karoubi}.}.

A unitary ghost free gauge fixing of the type described in section one for
a $U(N)$ gauge theory defines in a natural way an automorphism of the corresponding
vector bundle. In fact this type of gauge fixings associate in a unique way a gauge
transformation to each point of space-time. This is specially neat in the simplest
abelian Higgs model where the gauge condition $Im(\phi) = 0$ fixes in each
space-time point
a particular $U(1)$ rotation, namely that one killing the imaginary part of $\phi$.
Thus we can associate with these gauge fixings the same type of elements
used in the definition of $K^{-1}$. In fact for a $U(N)$ gauge theory once
we fix the non abelian part by impossing $X$ to be diagonal, we define the
automorphism
\be
\alpha(x) = e^{iX(x)}
\ee 
The charge in $K^{-1}$ is determined by the homotopy class of $\alpha(x)$.
Therefore what
this K-group classifies for gauge theories is, as expected, 
the topology of the gauge fixing.

In string theory the concept of K-theory as a way to describe RR charges is associated
with the physical picture of stability with respect to the processes of creation and
anhilation of D branes. In the case of gauge theories we have also a concept
of stability underlying the nature of the topology of gauge conditions. Namely if
we pass from a $U(N)$ to a $U(M)$ gauge theory with $M > N$ the very meaning
of abelian projection makes the charge of a magnetic monopole associated 
with the subgroup $U(N)$
completly independent of $M$, i.e stable under the changes of $N$ that correspond
in the gauge theory context to the type IIA process of creation anhilation
of ``elementary'' D-9 filling branes \cite{horova}. Moreover the topology survives in
the large $N$ limit.

An important aspect of $K^{-1}(X)$ is to be a group i.e to have defined the inverse.
In formal terms the inverse of the element $( I^{n}, \alpha)$ is $(I^{n}, {\alpha}^{-1})$.The reason
for this is that there exist an homotopy \cite{atiya} in $Gl(n,C)$ transforming continously
the matrix 
\be
\left(
\begin{array}{cc}
A&~\\
~&B\\
\end{array}
\right)
\ee
into the matrix
\be
\left(
\begin{array}{cc}
AB&~\\
~&1\\
\end{array}
\right)
\ee
In terms of type IIA branes 
what this operation represents is the following process. Take an unstable nine filling brane
with gauge bundle $E$ carrying some lower dimensional D brane charge. Now add another
nine filling brane with gauge bundle $E$ but carrying the opposite lower dimensional
D-brane charges. The net effect is the annihilation of the lower dimensional 
D-brane charges. In the case of gauge theories the equivalent process is just monopole
antimonopole annihilation.

\section{Dyons and M-Theory}
In type IIA string theory the group $K^{-1}$ was first interpreted from the point 
of view of M-theory.
A natural question is of course if the topology of gauge conditions in four
dimensional gauge theories which is also related to the $K^{-1}$ group is indicating some
type of M-theory nature of four dimensional gauge theories. The simplest answer is
that the relation of the gauge fixing magnetic monopole singularities with $K^{-1}$ is
related to the well known fact that these magnetic monopoles can be electrically charged,
becoming dyons. In fact, as first observed by Witten in \cite{dyon}, the electric
charge of monopoles in the presence of a non vanishing $\theta$ angle is due to the fact
that in the background of the monopole the gauge transformation defined by $e^{2\pi i Q}$,
where $Q$ is the electric charge generator, have non trivial winding number in the
$\pi_{3}$ of the gauge group. Now we easily identify this homotopy group with the one
associated with $\tilde{K}(S^{4})$ that, as we have discussed in section two, can be identified
with $K^{-1}(B^{3},S^{2})$ which is the relevant group 
for classifying the magnetic charge of the
monopole. 

A more direct way to understand the connection between $K^{-1}(X)$ and the electric charge
of the dyon is in terms of the moduli of a BPS monopole \cite{atiyah}. As it is well known
the moduli of one monopole is $M_{1} = R^{3} \times S^{1}$ where $R^{3}$ correspond to
the position of the monopole and $S^{1}$ arises from the gauge transformation
\be\label{ga}
e^{i \omega X}
\ee
on any solution. The generator $X$ is the field defined by (\ref{1}) used to fix the gauge.
In the moduli
space approximation \cite{manton} where we consider the effective lagrangian
describing the motion on the moduli, the generator of electric charge is precisely
$\partial_{\omega}$ for $\omega$ the $S^{1}$ angle. It is of course tempting
to associate
the necessity to work with $K^{-1}$ to the structure 
of the monopole moduli $R^{3} \times S^{1}$.
The piece $R^{3}$ can be identified with the codimension of the singularity and 
the electric phase in $S^{1}$ with the ``M-theory'' $S^{1}$.

In gauge theories this extra $S^{1}$ is not misterious at all, 
simply reflects the fact that the gauge fixing (\ref{1}) leaves 
unbroken the maximal abelian subgroup
and that $U(1)$ gauge transformations $e^{2\pi i X(x)}$ are non trivial, with 
non vanishing winding number on the monopole background.
The previous discussion leads to conjecture that the relevant
group to describe magnetic monopole singularities is $\tilde{K}(S^{3} \times S^{1})$,
where we compactify $R^{3}$ to $S^{3}$. Namely the K-group 
of the compactified magnetic monopole
moduli. From the M-theory point of view it seems that we promote the non 
trivial gauge transformations associated with the $S^{1}$ piece of $M_{1}$ 
to a physical extra dimension.

\section{Some Remarks and Speculations}
The goal in reference \cite{thooft} was to define a framework and a set of rules
for proving quark confinement. In order to have only abelian degrees of freedom and 
magnetic monopoles a good starting point is to use Dirac duality for
characterizing the confinement phase in terms of condensation of these magnetic monopoles. To
prove that such a condensation takes place is a hard open problem that requires to know
the dynamics of these monopoles and not just the topology that proves its existence. In
type IIA
string theory the claim is that for instance a D-6 brane is a bound state of two
unstable nine filling branes. In this case it is more or less clear that the vacuum of
the theory evolves from the unstable state represented by the filling nine branes into
the state with stable six branes. The equivalent process in gauge theories,
where we replace the six brane by the magnetic monopoles associated with the topology
of the gauge fixing, would strongly indicate permanent quark confinement i.e condensation of
these monopoles into the vacuum. What we are missing in gauge theories is the analog
of the unstability of the nine filling brane in type IIA, although the minus 
sign of the beta function
is probably an indirect way to point out to this unstability.

One type of D-brane we have not discussed is the 8-brane in type IIA. 
This brane is difficult
to characterize in terms of gauge fixing singularities. Its K-theory 
description would be given
in terms of $K^{-1}(S^{1})$ and require
some $N= \infty$ limit of a discrete $Z(N)$ with $N$ the rank of the gauge group. 
In four dimensional gauge theories such a type of brane
would correspond to a domain wall with ``RR field strenth'' 
$\hbox{Tr} F \wedge F$ i.e the topological
charge \footnote{ Here we are using the relation between RR field strength and the
gauge connection of a $U(N)$ bundle in type IIA suggested in 
reference \cite{mw},\cite{witten2}. 
In this simbolic sense we can talk of RR field forms in a gauge theory.}. It would
be possible that QCD strings would appear as strings ending on 
these domain walls, in the
same spirit that in MQCD.

If we are bold enough to follow the analogy with M-theory, we should consider
the magnetic monopoles associated with the singularities of the gauge fixing as the
analog of the D-0 branes in type IIA and the corresponding 
Matrix model ( dimensional reduction
to $0+1$ of four dimensional gauge theory ) as the dynamics
of these magnetic monopoles. Following this way of thinking you could expect
that, similarly to what happens in Matrix theory, an extra dimension
( in the gauge case a five dimension ), should show up dynamically in
processes of monopole scattering, very much in the same way as the eleven dimension
appears by comparing the scattering of D-0 branes in Matrix theory with the
eleven dimensional supergravity result \cite{bb}. This gravitational description,
if it exist,
would be very much at the core of the holographic approach.

\section*{Acknowledgments}
I thank E.Alvarez for many fruitful discussions.
This work has been partially supported by the
the Spanish grant  AEN2000-1584.



\end{document}